\documentstyle{elsart}

\begin{document}
\begin{frontmatter}

%\bibliographystyle{unsrt}

%\draft
%\widetext
\title{Experimental studies of T--shaped quantum dot transistors:
phase-coherent electron transport} 
\author[Cav]{C.--T.~Liang},
\author[Cav,Shef]{J.~E.~F.~Frost},
\author[Cav]{M.~Pepper},
\author[Cav]{D.~A.~Ritchie} and
\author[Cav]{G.~A.~C.~Jones}

\address[Cav]{
Cavendish Laboratory,
Madingley Road,
Cambridge CB3 0HE,
United Kingdom}

\address[Shef]{Department of Electronic and Electrical
Engineering, University of Sheffield, Mappin Street, Sheffield S1 3JD,
United Kingdom}

\date{\today}

\maketitle

%\widetext
\begin{abstract} 
%\leftskip 54.8pt
%\rightskip 54.8pt

We have measured the low-temperature transport properties of a T-shaped
quantum dot. Replicated oscillations superimposed on one-dimensional 
conductance steps are observed. These structures are consistent with
electron phase-coherent length resonance effects in the ballistic
regime. Using a simple model, we suggest that both one-dimensional and
two-dimensional electron-electron scattering gives rise to 
electron phase-breaking in our system. 

\end{abstract}


\end{frontmatter}
\newpage
 
%\section{Introduction}
The ballistic transport of electrons confined in one-dimensional (1D) 
systems has been extensively studied both experimentally and
theoretically. Conductance quantisation observed in split-gate devices
\cite{Wharam,vanwees} provides clear evidence for ballistic
transport. The interference features observed in a split-gate device
\cite{Brown} and in double-slit experiments \cite{Yacoby} 
have clearly demonstrated the
existence of phase coherent-electron transport in the
ballistic regime. For a ballistic channel patterned with
barriers, there are predictions of electron phase-coherent length 
resonance effects \cite{Lent,Wu} in such a structure. Therefore
oscillations superimposed on ballistic conductance steps may be
observed in this regime as has been recently reported when there is
only one 1D subband occupied \cite{Frost}.

We have now extended our previous work and in this Communication, we 
present experimental studies of a T--shaped quantum dot transistor.
We observe replicated conductance oscillations superimposed on 1D
conductance steps which are interpreted as electron phase-coherent
length resonance effects in the ballistic channel. The magnetic field
and temperature dependence of these structures will be described. Four
samples showed similar characteristics, and measurements taken from
one of these are presented in this paper.

The Schottky gate pattern shown in the inset to Fig.~1
was defined by electron beam lithography on the surface of a
GaAs/Al$_{0.3}$Ga$_{0.7}$As heterostructure, 100~nm above a two-dimensional
electron gas (2DEG). The carrier concentration of the 2DEG is
$4.7\times 10^{15}$ m$^{-2}$ with a mobility of 150 m$^{2}$/Vs.
Experiments were performed in a $^3$He cryostat at 300 mK and
the two-terminal differential conductance $G=dI/dV$ was measured using
an ac excitation voltage of 10~$\mu$V.

Patterning of the underlying 2DEG most closely mirrors the shape of
the lithographically defined metallisation at the gate voltage when
depletion just occurs under the 100~nm width fingers. This is at a
higher voltage than that needed to deplete carriers from beneath 
the wider metallisation due to fringing effects. Therefore the
maximum potential modulation along the channel is obtained when one
side of the device is maintained at a constant voltage and the other
is swept to reduce the device conductance \cite{Frost,Tomas}. In our
case, experiments were performed with one side of the device at a
fixed voltage and the other gate is swept. Figure~1 shows $G(V_{g1})$
for a fixed $V_{g2}$ in two different cooldowns. In both cases we
observe similar structures, in particular two conductance dips
superimposed on two 1D conductance steps, suggesting that
impurity scattering is not significant in our system, and the
oscillating features primarily arise from the potential landscape
in the channel defined by the lithographic pattern. We interpret 
these conductance oscillations as electron phase-coherent length
resonance effects between two barriers.

Figure~2 shows $G(V_{g1})$ for $V_{g2} = -1.4$~V at various magnetic
fields $B$. The oscillations superimposed on the conductance steps 
gradually disappear as $B$ is increased. The application of a
perpendicular magnetic field breaks time reversal symmetry, suppressing
coherent backscattering of electrons in the ballistic 
channel. This is analogous to suppression of weak localisation 
effects in diffusive transport by a perpendicular magnetic field.
           
Figure~3 shows $G(V_{g1})$ for $V_{g2} = -1.4$~V at different
temperatures $T$. As $T$ is increased, the conductance oscillations
become weaker and we recover ballistic conductance steps.
This effect is due to a reduction in phase coherence time at higher 
temperatures, leading to an enhancement of phase-breaking rate. 
Now we use a simple model to determine the electron
phase-breaking mechanism in our system. Provided that the amplitude
between a peak and a trough in conductance oscillations $Amp$ is given
by

\begin{equation}
Amp\propto\mathrm{exp}(-\tau/\tau_{\phi}),
\end{equation}

where $\tau$ and $\tau_{\phi}$ are the traversal time for electron to
pass through the ballistic channel and electron phase coherence time,
respectively. From this we know that

\begin{equation}
\mathrm{ln}({\em Amp \/})\propto-\tau/\tau_{\phi} + {\em A \/} ,
\end{equation}

where $A$ is a constant. Here we assume that the traversal time $\tau$
is temperature independent which holds when the thermal broadening
is much smaller than the Fermi energy in our case. From the amplitude
of the conductance oscillations at various temperatures, we have

\begin{equation}
\mathrm{ln}({\em Amp(T) \/})\propto -1/\tau_{\phi}({\em T\/}) + {\em A\/}/\tau.
\end{equation}

Figure~4 shows the normalised logarithm of the conductance 
oscillations amplitude on the second step (marked by circles) and on the
first step (marked by squares) as a function of temperature.
From the fit we obtain 1/$\tau_{\phi}(T) = 
0.2147~T+0.0717T^{2}$ln($T$). There are two physical mechanisms which
can give rise to the linear term in $T$: 2D electron-electron
scattering  \cite{Uren2,Davies,Uren3},
enhanced by disorder 
in the diffusive limit \cite{Pooke1,Pooke2}, and
electron-electron scattering in the clean metallic 
regime where only a small number of 1D subbands 
are occupied \cite{Luttinger}.
In our system, at elevated temperatures we observe 
well quantised conductance steps,
demonstrating that our device is in
the ballistic rather than the diffusive limit.
Moreover, the product $k_{F}${\em l\/} in the bulk is $\approx$3000 
where $k_{F}$ is the Fermi wave-vector and {\em l\/} is the elastic
scattering length, respectively, indicating that the device is in the 
clean metallic regime. Combining these two facts,     
we conclude that the linear term in $T$ corresponds to
electron-electron scattering in 1D in the clean metallic 
limit \cite{Luttinger} whereas 
the $T^{2}$ln($T$) term corresponds to the well-known electron-electron
scattering in 2D \cite{Chaplik,Quinn}. The results suggest that
both 1D and 2D electron-electron scattering are important in our
system: 2D scattering occurs in the wide regions of the channel
where there are many subbands occupied, whereas 1D electron
scattering takes place around two barriers, the narrow regions of the
channel where only one or two subbands are occupied.

In conclusion, we have measured the low-temperature transport
properties of a T--shaped quantum dot transistor. We have observed
replicated oscillations superimposed on ballistic conductance 
plateaux which are interpreted as electron phase-coherent
length resonance effects in the ballistic channel. Using a simple
model, we suggest that electron-electron scattering both in one
and two dimensions introduces electron phase-breaking in our system. 
As we have well characterised T--shaped quantum
dot devices, in future this work can be extended by
investigating a finite-period T-shaped dot array \cite{Leo}
where formation of energy gaps arising from Bragg
reflections in an artificial lattice is expected to be
observed in {\em zero\/} magnetic field.

This work was funded by the UK
Engineering and Physical Sciences Research Council, and in part
by the US Army Research Office. We thank C.H.W. Barnes 
 and C.G. Smith for helpful discussions, N.P.R. Hill for experimental
assistance, and H.D. Clark for his help in sample preparation. DAR 
acknowledges support from the Toshiba Cambridge Research Centre.

\newpage

%\begin{references}

\newpage

\centerline{\bf Figure Captions}

Figure 1.
%\begin{figure}
%\epsfxsize=70truemm
%\centerline{\epsffile{jeff1.ps}}
%\caption {
Lower curve: 
differential conductance measurements as a function of gate
voltage $G(V_{g1})$ for $V_{g2}= -1.4$~V in the first cooldown.
Higher curve (offset by 50~$\mu$S for clarity): $G(V_{g1})$ for
$V_{g2} = -1.72$~V in the second cooldown.
The inset shows the device geometry.
%} \label{1} \end{figure}

Figure 2.
%\begin{figure}
%\epsfxsize=70truemm
%\centerline{\epsffile{jeff2.ps}}
% \caption { 
$G(V_{g1})$ at various magnetic fields $B$ for $V_{g2} = -1.4$~V. From
bottom to top: $B = 0$ to $1$~T in 0.1~T steps. Traces are
successively offset by 30~$\mu$S for clarity.
%} \label{2} \end{figure}

Figure 3.
%\begin{figure} 
%\epsfxsize=80truemm 
%\centerline{\epsffile{jeff3.ps}}
% \caption {
$G(V_{g1})$ at different temperatures for $V_{g2} = -1.4$~V. From
bottom to top: $T= 0.3$~K to 1.5~K in 0.1~K steps and 
$T = 1.85$~K to 3.25~K in 0.1~K steps. Curves are offset successively
by 15$\mu$S for clarity. 
%} 
%\label{3} \end{figure}

Figure 4.
%\begin{figure} 
%\epsfxsize=80truemm 
%\centerline{\epsffile{jeff4.ps}}
% \caption {
Normalised logarithm of the amplitude $Amp$ of the conductance
oscillations on the first conductance step (marked by squares) and on the
second step (marked by circles) at various temperatures $T$. The solid
line shows a fit ln$(Amp(T))= -1.09 -0.2147$~$T-0.0717~T^{2}$ln($T$).
%} 
%\label{4} \end{figure}

%\end{multicols}
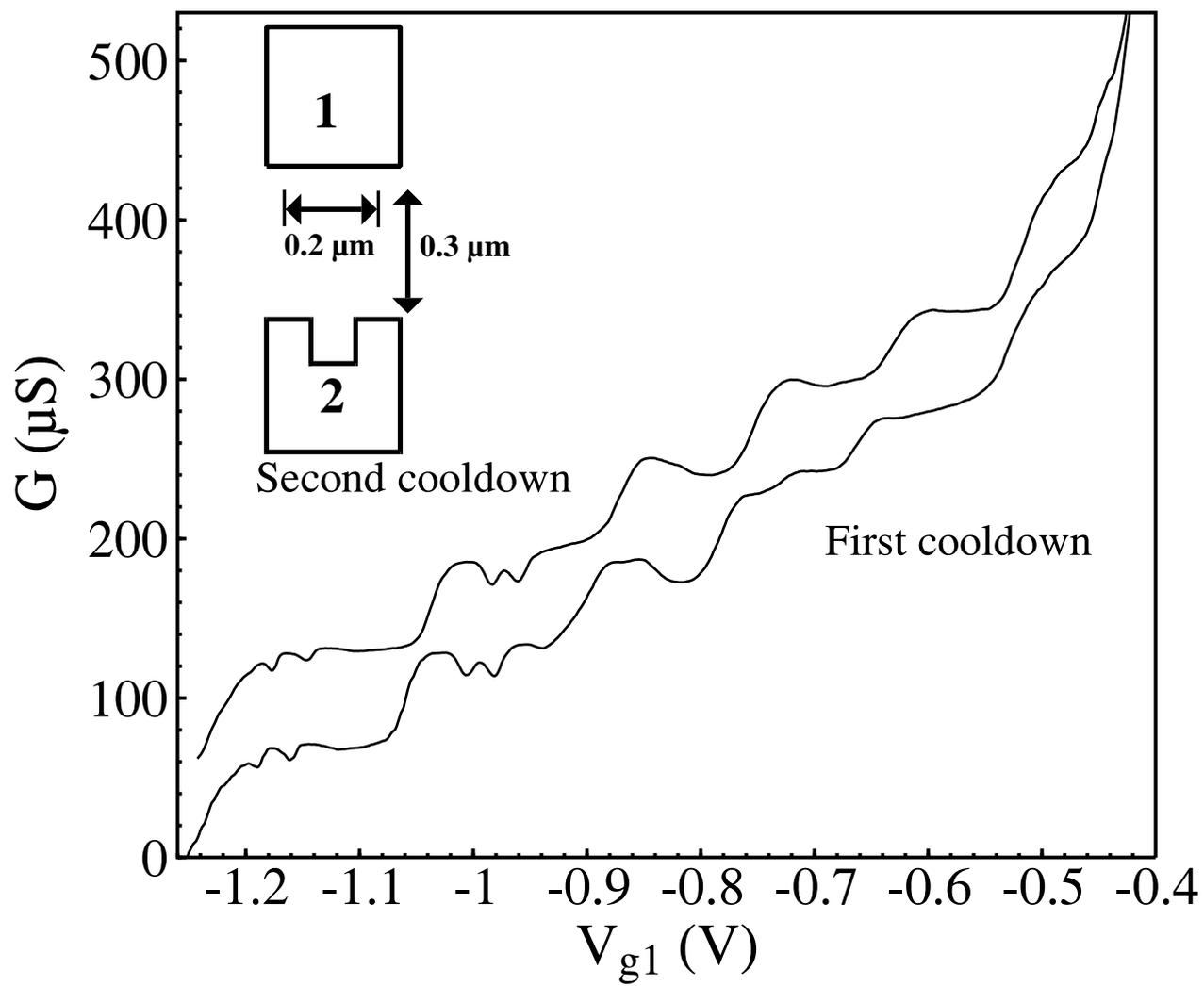
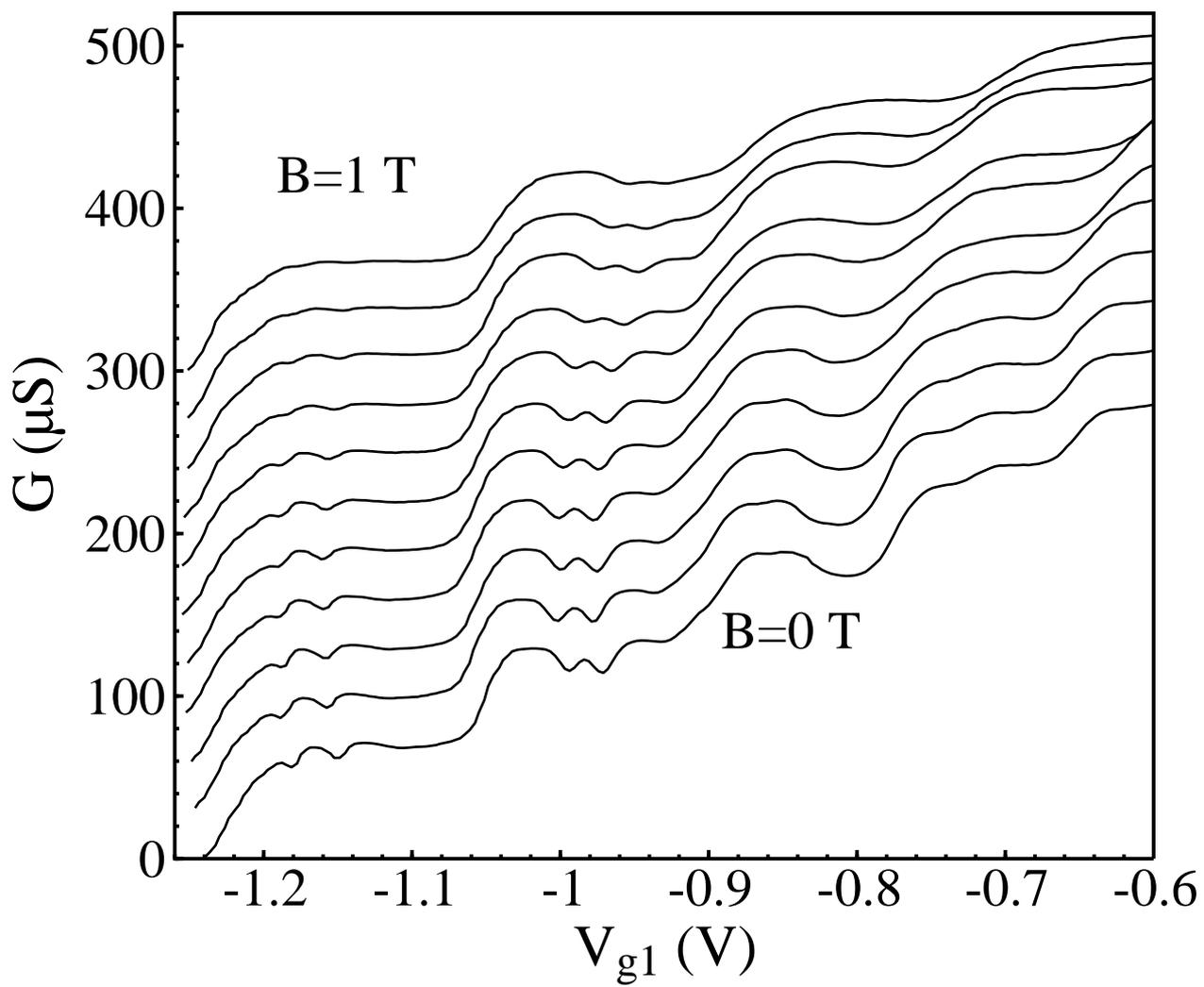
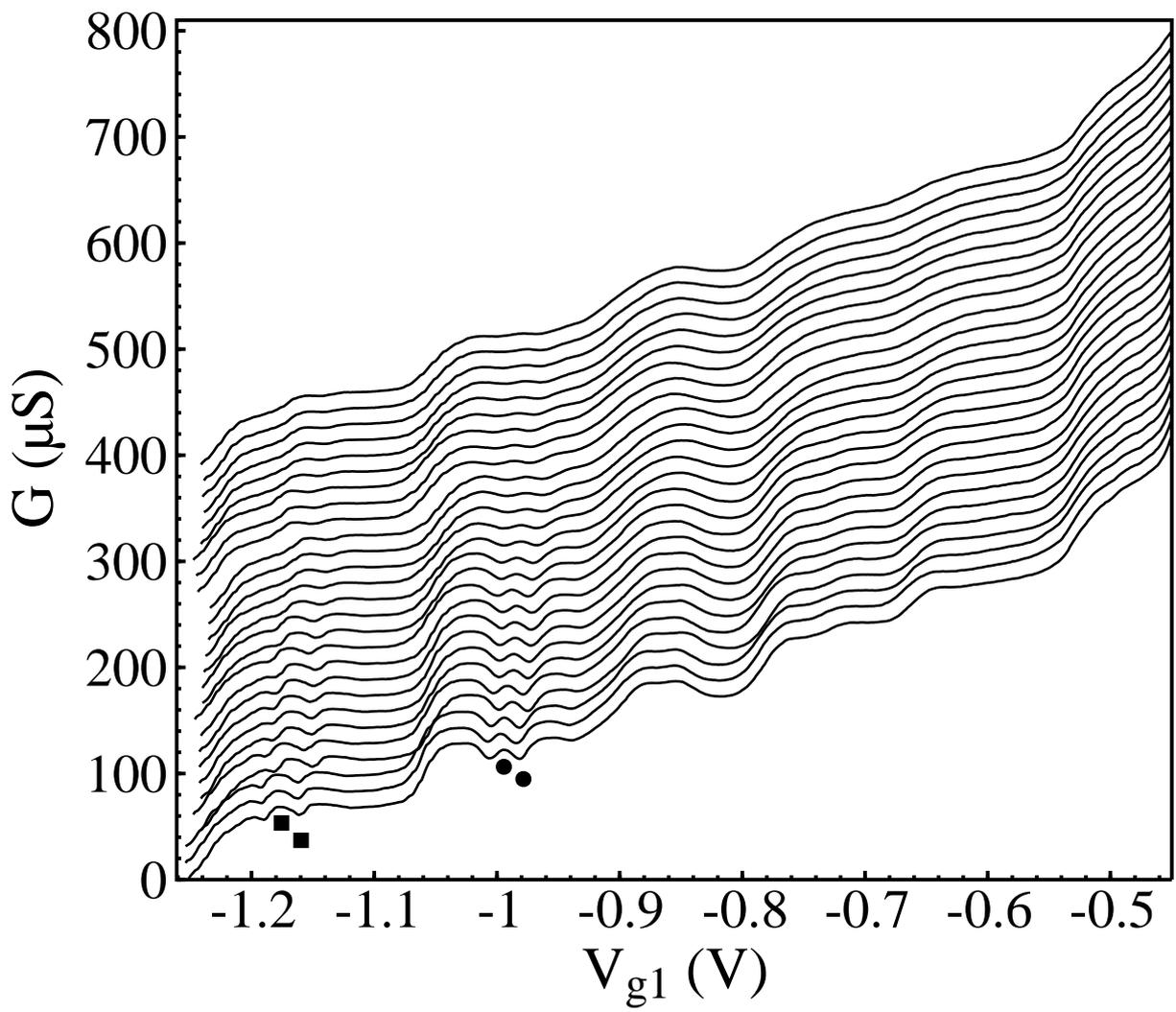
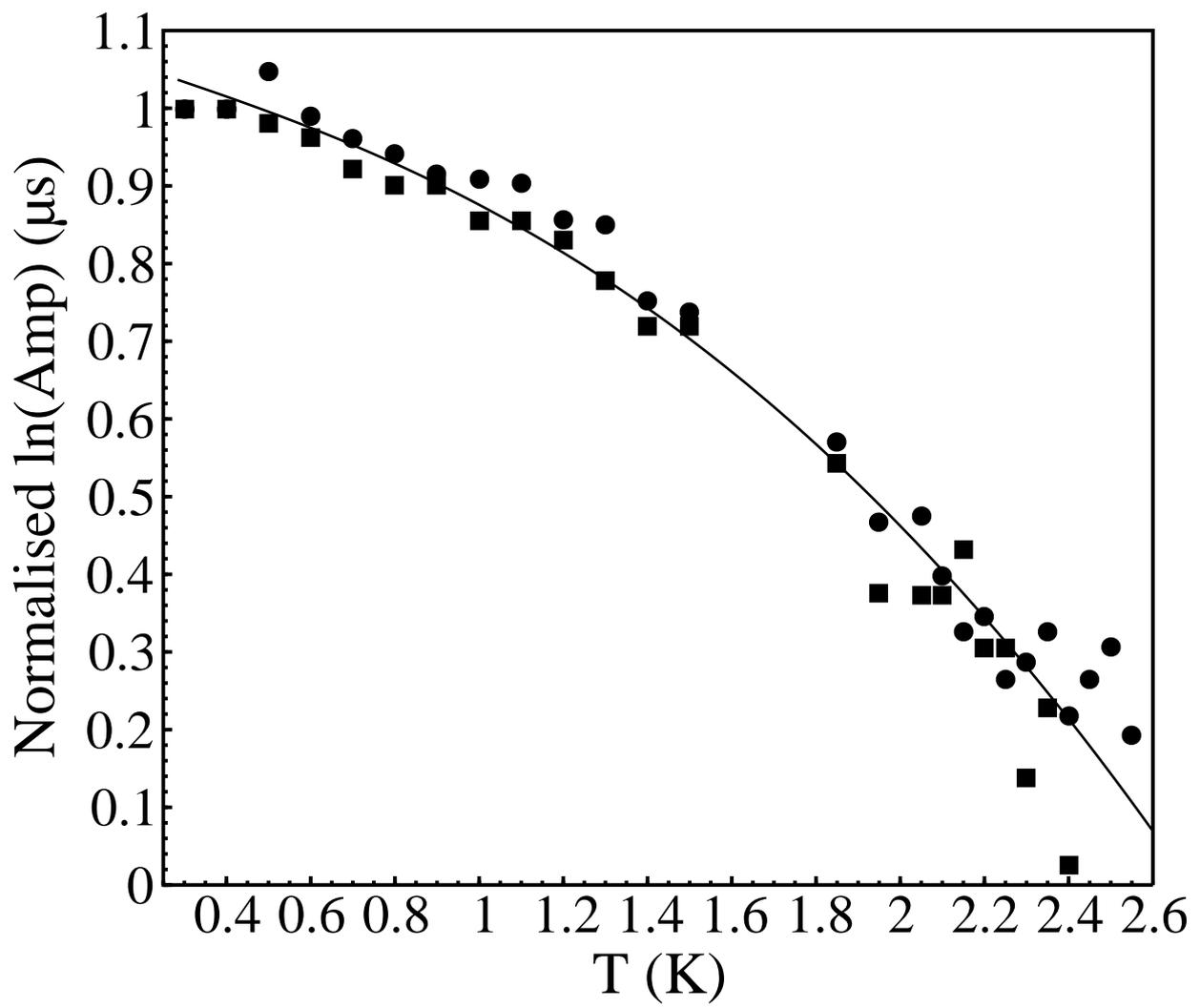
\end{document}